\documentclass[preprint,proceedings]{rmaa}
\usepackage{paralist}
\usepackage{psfrag,color}

\def\Msol{\thinspace\hbox{$\hbox{M}_{\odot}$}}

\SetYear{2003}

\title{The intrinsic high metallicity of cSNRs}

\author{A. Rodr\'{\i}guez-Gonz\'alez \altaffilmark{1}, 
        G. Tenorio-Tagle  \altaffilmark{1},
        S. Silich \altaffilmark{1}}

\altaffiltext{1}{Instituto Nacional de Astrof\'\i sica Optica y Electr\'onica, 
AP 51, 72000 Puebla, M\'exico}

\shortauthor{Rodr\'{\i}guez-Gonz\'alez}
\shorttitle{cSNRs}

\listofauthors{A. Rodr\'\i{}guez-Gonz\'alez, G. Tenorio-Tagle, \& S. Silich}
\indexauthor{Rodr\'\i{}guez-Gonz\'alez, A.}
\indexauthor{Tenorio-Tagle, G.}
\indexauthor{Silich, S.}

\abstract{Here we present semi-analytical solutions of the hydrodynamics of 
compact supernova remnants (cSNRs) and their chemical evolution. Two
models of $15 \; M_{\odot}$ and $25 \; M_{\odot}$ supernova explosions 
evolving in a high density medium ($n_0 \sim 10^7 \; cm^{-3}$) are 
throughly discussed. We demostrate, that after a few years of
evolution, diffusion leads to an homogeneous distribution of the SN 
products inside the cold, thin shell formed after cooling of the 
swept-up matter sets in. We find metallicity values within a range 
$5.5 \; \le \,Z_O/Z_{\odot} \, \le \; 6.0$ for $M_{SN}= 25\; M_{\odot}$
and $1.0 \; \le \,Z_{O}/Z_{\odot} \, \le \; 1.5$ for $M_{SN}= 15\; M_{\odot}$,
and speculate wheather these metallicities could be 
related to the observed QSOs super solar elemental abundances
(Ferland et al. 1996, Hamann \& Ferland 1999). }

\resumen{Aqu\'\i{} presentamos soluciones semianal\'\i{}ticas de la 
hidrodin\'amica de remanentes de supernova compactos (cSNRs) y su 
evoluci\'on qu\'\i{}mica. Discutimos dos modelos de explosiones de 
supernova de $15 \; M_{\odot}$ y $25 \; M_{\odot}$, envueltas en un 
medio denso  ($n_0 \sim 10^7 \; cm^{-3}$). Demostramos que despu\'es 
de algunos a\~nos de evoluci\'on, la difusi\'on produce una
distribuci\'on homog\'enea de los productos de la SN dentro de
los delgados cascarones fr\'\i{}os formados despu\'es del enfriamiento 
del material barrido. Encontramos metalicidades en un intervalo
de $5.5 \; \le \,Z_O/Z_{\odot} \, \le \; 6.0$ para $M_{SN}= 25\; M_{\odot}$
y $1.0 \; \le \,Z_{O}/Z_{\odot} \, \le \; 1.5$ para $M_{SN}= 15\; M_{\odot}$,
y especulamos que estas metalicidades est\'an relacionadas con las abundancias
observadas en QSOs (Ferland et al. 1996, Hamann \& Ferland 1999).}

\addkeyword{Compact Supernova Remnants}
\addkeyword{Quasars}
\addkeyword{High metallicities}

\begin{document}
\maketitle

\section{Introduction}
\label{sec:intro}

The discovery of the high-redshift quasars (QSO) implies
that massive star formation occurs very early in galaxy evolution. 
Spectra of high redshifted QSOs ($z\,>\,4$) present broad line
emission (Hamann \& Freland, 1993) which is similar to that
seeing in active galactic nuclei (AGNs). 
The metallicities of the associated fast moving gas derived from 
the spectral analysis are surprisingly high for so young objects 
and fall into the range $Z_{\odot} \le Z \le 9Z_{\odot}$ 
(Hamann \& Ferland 1999). 
These results have a great interest because provide information about
an early history of star formation in galaxies. If high metallicity
is related to a short episode of star formation, it has to occur
at early (few Gyr) stages of galaxies evolution and suggests a fast
mixing of ejected metals with a bulk of primordial low metalicity
gas (Hamman \& Ferland, 1993; Ferland et al. 1996). Different
modifications of this scenario have been proposed by Silk \& Rees (1998) 
who have discussed an inhenced star 
formation inside an accretion disk.

The observed quasar (and AGN) spectra show a great similarity to the
spectra of type II peculiar supernova remnants (Terlevich et al. 1992) 
which one believes to be a compact SNR (cSNR) evolving in a very dense 
($\sim 10^7$ cm$^{-3}$) circumstellar environment. Therefore here we 
analyze an alternative scenario based on the starburst model 
of AGNs (Terlevich et al. 1992). For the calculations we use
a semi-analitycal model of the young supernova remnant evolution 
and thoroughly discuss the dispersal of the ejected metals into
the swept-up primordial gas. We simulate cSNRs evolution from the
15 $M_{\odot}$ and 25 $M_{\odot}$ progenitors and show that model
predicted thicknesses, densities and temperatures allow a fast mixing
of the abundant ejecta with the swept-up low metallicity circumstellar
material. The resultant cSNR abundances fall into the observed
QSOs metallicity range.

\section{The input model}
\label{sec:met}

\subsection{Ejecta model}

The interaction of the SN ejected matter with the surrounding medium 
causes a four zone structure: the inner free expanding ejecta which is 
bounded by a reverse shock with radius R$_{RS}$ and velocity 
D$_{RS}$; the shocked ejecta which is accommodated between the 
reverse shock position and the contact discontinuity R$_{CD}$;
the shocked CSM which is separated from the ejected material by 
the CD and from the surrounding medium by the leading shock with 
radius R$_{LS}$ and velocity D$_{LS}$. 
The outward shock sweeps-up the dense surrounding medium and enhances its 
temperature to $\sim 10^8$K. The reverse shock is slower and
thermalizes the SN ejecta. 

The cSNR initial evolution is dependent on the surrounding gas number 
density and the ejecta density and velocity radial distribution. 
We adopt Franco et al. (1991) analytical model based on Woosley et
al. (1988) calculations. It assumes that supernova ejecta obey the
free expansion law with $r^{-3}$ density and linear velocity
distributions:
\begin{equation}
\hspace{0.0cm} \rho_{ej}(r,t) = \left\{
\begin{array}{lcl}
M_{ej}\,r^{-3}/4 \pi \ln{\left(\frac{R(t)}{R_c}\right )}, 
\hspace{0.4cm} r \ge R_c \\[0.05cm]
\hspace{1.0cm} 0 \hspace{2.6cm} r < R_c\\
\end{array}
\right .
\end{equation}
\begin{equation}
\hspace{-0.0cm} V_{ej}(r,t) = \left\{
\begin{array}{lcl}
\frac{r-R_c}{R(t)-R_c}V_m, \hspace{1.95cm} r \ge R_c\\[0.05cm]

\hspace{1.0cm}0 \hspace{2.56cm} r < R_c\\
\end{array}
\right . 
\end{equation}
where $R(t) = R_m + V_mt$, $R_m$ is the initial radius and $V_m$ 
is the maximum ejecta velocity, $R_c$ is the inner radius of the
ejecta. The total mass of ejecta and masses of the ejected oxygen 
and iron have been taken form Woosley (1978) model. The initial model 
parameters are summarized in the Table 1.

\begin{table}[!hbp]
\begin{center}
\caption{Initial Conditions}
\begin{tabular}{lccc}
\hline
\hline
& $15 \; M_{\odot}$ & $25 \; M_{\odot}$ \\
\hline
\hline
Initial energy [$erg$]& $1\times10^{51}$&$1\times10^{51}$ \\
\hline
Ejecta mass [$M_{\odot}$] & $13.44$& $23.61$\\
\hline
Maximum & $6000$& $5000$\\
velocity [$km \; s^{-1}$] \\
\hline
$R_c$ [$cm$]&  $6.45 \times 10^{15}$&  $3.24 \times 10^{15}$\\
\hline
$R_{m}$ [$cm$]&  $3.82 \times 10^{16}$&  $4.10 \times 10^{16}$\\
\hline
Ejecta mass & $0.42 $& $3.0$\\
of oxygen [$M_{\odot}$]\\
\hline
Ejecta mass & $0.06$& $0.08$\\
of iron [$M_{\odot}$]\\
\hline \hline
\end{tabular}
\begin{list}{}{}
\item $^{\mathrm{}}$ 
\end{list}
\end{center}
\end{table}

\subsection{The dispersal of metals}

There are essentially two irreversible mass transport mechanisms
leading to the dispersal of the heavy elements released during the
supernova explosions throughout the ISM. They are diffusion and
turbulent mixing (Roy \& Kunth, 1995; Tenorio-Tagle, 1996, Oey 2003).
The determination of the appropriate turbulent parameters so far is speculative.
Therefore here we focus on the diffusion of the supernova ejected
metals and their mixing with the swept-up CSM. 

We assume that the ejecta represents a homogeneous mixture 
of the progenitor outer hydrogen envelope with the synthesized heavy 
elements. At the contact discontinuity this high
metallicity gas is mixed with the low metallicity circumstellar 
material. The characteristic diffusion time over a length L is 
$t_D = L^2 / D$, where D is the diffusion coefficient. 
When the dynamical time t exceeds the characteristic diffusion time t$_D$, 
the metallicity of the mixture approach the value
\begin{equation}
      \label{eq.10}
Z_A(t)/Z_{\odot}=\frac{M_{A\;ej}(t)/Z_{A\;\odot}+Z_{ISM}\,M_{ISM}(t)}
{M_{ISM}(t)+\chi M_{ej}(t)}
\end{equation}
where $M_{CSM}(t)$ and $\chi M_{ej}(t)$ are masses of the swept-up CSM 
and the ejecta, respectively, and  $M_{A\;ej}(t)$ is the mass of the 
element A that has been accumulated within a given gaseous segment.

\section{The results of the calculations}

\subsection{Hydrodynamical evolution of the cSNR}

Figure 1 shows runs of leading and reverse shocks velocities and radii 
for 15\Msol supernova remnant. The leading shock has initial speed 
$\sim 4200$ km s$^{-1}$. Strong radiative cooling in the swept-up CSM 
matter induces a sudden drop in the post-shock thermal pressure and a
noticeable deceleration of the leading shock velocity which falls 
down from $\sim 3000$ km s$^{-1}$ to $\sim 2000$ km s$^{-1}$ during 
a short transition phase. The process is followed by the condensation 
of the swept-up interstellar gas into a thin dense shell. The shell
density exceeds the shocked ejecta density by about four orders of 
magnitude. Therefore the CD adjusts the leading shock position and later 
on follows its expansion. Radiative phase starts at t$_{sg} \sim
1.2$yr after explosion and is well identified as breaks of the solid 
lines at velocity curve. 
\begin{figure}[!hbpt]
\vspace{5.4cm}

\includegraphics{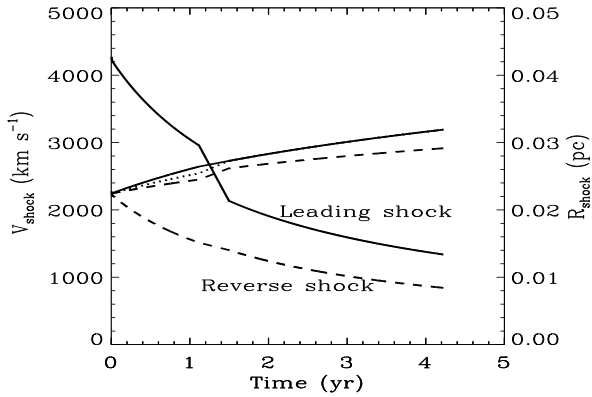}

\caption{Radius and velocity evolution of the shocks.}
\label{fig1}
\end{figure}
The reverse shock is substantially slower. It remains adiabatic for a 
much longer time because of the smaller post-shock density and thus 
less effective cooling. Shock waves heat both the swept-up
circumstellar material and the shocked ejecta to high temperatures 
$\sim 10^8$K. However the temperature of the shocked interstellar
material suddenly drops when cooling sets in and later on 
is supported at $10^4$K level by the ionizing photons arising from the
cooling shocked gas.

Figure 2 shows the swept-up mass time evolution. The interstellar mass
accumulated in the outer shell grows almost linearly with time and at 
the end of the calculations reaches around 10\Msol. During this time
the reverse shock processes only $\sim 4$\Msol of the ejecta which
is spread inside a relatively thick inner shell. The flat parts of
the curves in Figure 2 come from our simplified description of
the swept-up interstellar mass condensation when catastrophic 
radiative cooling sets in.
\begin{figure}[!hbpt]
\vspace{5.3cm}
\includegraphics{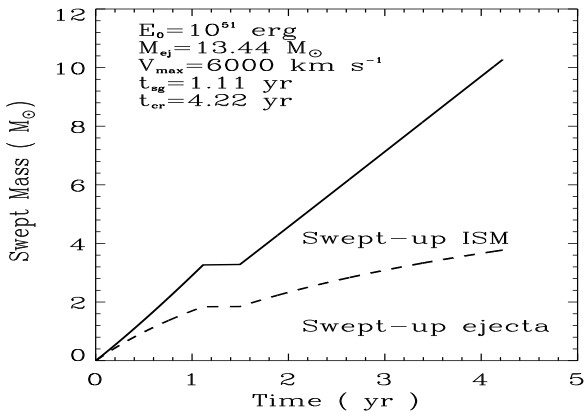}
\caption{Swept-up mass evolution.}
\label{fig2}
\end{figure}

\subsection{Chemical evolution of the cSNR}

Using the shell thickness and temperature given in the previous section 
we calculate the characteristic diffusion time t$_d$ one needs to
mix the ejected metals with the swept-up CSM. At the beginning of the
calculations the characteristic diffusion time grows with time due to 
the growth of the shell thickness. 
However when radiative cooling sets in it drops about four orders of 
magnitude because of the sudden decrease of the outer shell
thickness. This leads to an effective and rapid mixing of the ejected 
metals with the swept-up interstellar matter and defines the fast 
chemical evolution of the remnant.
Initial metallicities of the swept-up interstellar medium and the
ejecta and the resultant metallicity of the cSNR outer shell 
are shown in Figure 3. 
\begin{figure}[!hbpt]
\vspace{5.0cm}
\includegraphics{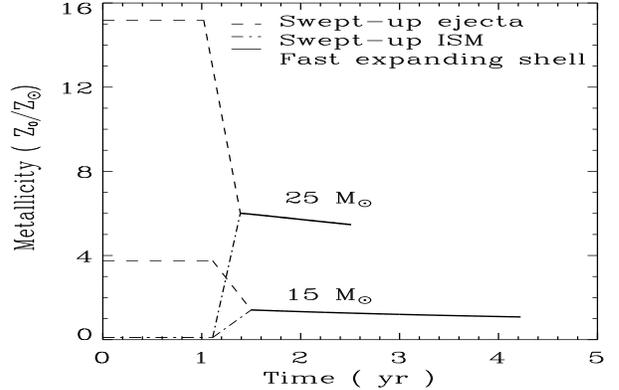}
\caption{The cSNR metallicity as function of time if oxygen is
used as a tracer.}
\label{fig4}
\end{figure}

\newpage
\section{Conclusions}
\label{sec:conc}

The calculations show that effective metal diffusion occurs at early
cSNR evolution, soon after a dense shell formation. The resultant
shell metallicity was found to be within the range
$1.0 \, \le \, Z/Z_{\odot} \, \le \, 6.0$ if oxygen is used as a tracer and 
$0.65 \, \le \, Z/Z_{\odot} \, \le \, 1.3$ if iron is used as a tracer of
the gas metallicity. This is in a remarkably good coincidence with the 
observed QSO metal abundances. 
We speculate therefore that the observed QSO metallicities could be 
related to SN explosions immersed in a high density CSM.\\

The authors feel honoured to have attended this meeting and 
acknowledge support from CONACYT grant 36132-E.

\end{document}